# Insights into the mechanical properties and fracture mechanism of Cadmium Telluride nanowire.

Md. Adnan Mahathir Munshi[1], Sourajit Majumder[1], Mohammad Motalab[1],*, Sourav Saha[1,2].

[1]Department of Mechanical Engineering, Bangladesh University of Engineering and Technology, Dhaka-1000, Bangladesh.

[2]Theoretical and Applied Mechanics, Northwestern University, Evanston, Illinois, USA.

**Abstract**

The mechanical properties of Cadmium telluride (*CdTe*) nanowire have become focus of interest now-a-days due to its promising application in opto-electro-mechanical nanodevices. In this study, molecular dynamics simulations have been used to investigate the mechanical behavior of Zinc Blende (ZB) crystal structured *CdTe* nanowires (NWs) by varying size, temperature, crystal orientation and strain rate under tension and compression. Results show that the fracture strength of the [111]-oriented *CdTe* NWs is always higher than that of the [110]-oriented *CdTe* NWs under tension whereas in compression, the fracture strength of the [111]-oriented *CdTe* NWs is significantly lower than that of the [110]-oriented *CdTe* NWs. Moreover, under tensile load, void in ZB [111]-oriented *CdTe* NWs has been observed which is a new failure mechanism found in this study. It has also been observed that size has negligible effect on the tensile behavior but in compression the behavior is clearly size dependent. Both tensile and compressive strengths show an inverse relation with temperature. When tensile load is applied along NWs growth direction, the [111]-oriented *CdTe* NWs fail by creating void in [10-1] direction regardless of temperature and NW size. Under compression, the [111]-oriented nanowire show buckling and plasticity. Finally, the impact of strain rate on [111]-oriented ZB *CdTe* NWs is also studied where higher fracture strengths and strains at a higher strain rates have been found under both tension and compression. With increasing the strain rate, the number of voids is also increased in the NWs. This study will help to design *CdTe* NWs based devices efficiently by presenting in-depth understanding of failure behavior of the [111]-oriented *CdTe* NWs.

# 1.Introduction

Semiconductor nanowires (NWs) are one of the most basic elements of future electro-mechanical devices. Directed growth of semiconductor nanowires has important effect on its different properties and has attracted scientific interest in recent years as an important promoter of nanotechnology. While mechanical properties of semiconductor nanowires are of immense importance as they work as building blocks for developing MEMS and NEMS systems. These properties can be utilized to produce effective and functional devices[1–4]. The recent rapid growing scientific and technological interest on *CdTe* arises from its special physical properties, i.e. its mechanical, chemical and thermal stability. Moreover, *CdTe* is a promising semiconductor material for electronic and optical devices [5–8]. As a group II-VI semiconductor material, *CdTe* has a direct band gap of 1.5 eV at room temperature[9], which makes it a perfect and ideal material candidate for high-efficiency solar cells [10]. While, *CdTe* NWs are promising material candidate for applications in high-performance photodetectors, field effect transistors and solar cells [11,12].

NWs are suitable for studying the fundamental deformation mechanisms of semiconductor materials. Meanwhile, the nanostructure materials are more suitable to carry the tensile and compressive load due to its enhanced mechanical properties than the traditional bulk counterpart[13–15]. For the purpose of designing and fabricating NW-based photo-mechanical devices it is important to describe and predict the mechanical behavior of different NWs. Mechanical properties of semiconducting NWs were studied by tensile tests[13,16,17], compression tests[18,19]and bending test[20–22]in many in situ studies. Recently, molecular dynamics (MD) simulations is widely used to investigate not only the mechanical behavior but also fracture mechanisms of various NWs along with experimental and in situ studies[23–25].Temperature, diameter, and strain rate-dependent mechanical behavior under tension for Si NWs were investigated successfully by Kang and Cai using molecular dynamics[26]. Remarkable phenomena like shear failure and cleavage failure of Si NWs were explained in that investigation. Tsuzuki et al.[27] showed the failure mechanism of SiC NWs under both tensile and compressive loading. Here, it was found that ZB SiC NWs displays complex plasticity before failure while wurtzite SiC NWs is brittle in nature. Furthermore, Cheng et al.[28] had reported the fracture strength due to the size dependency of SiC NWs. It was found by pial et al. that the direction of cleavage planes of ZB InP NWs change with temperature under tension while investigating the mechanical behavior of InP NWs[29]. The mechanical behavior of many other semiconductor NWs, such as Ge [30] , GaN[31] and ZnO[32] were also investigated successfully with MD simulations .

However, there is hardly any research on mechanical properties of *CdTe* NWs by simulations. Although the applicability of *CdTe* NWs requires in-depth knowledge of their mechanical properties and failure behavior, the number of experimental investigations is inadequate. Moreover, the failure behavior of *CdTe* NWs has yet to be discussed. Keeping this scope in mind, this paper presents atomistic simulation results of *CdTe* NWs under uniaxial tensile and compressive load. The effects of crystal orientation, temperature, size, and strain rate on the mechanical properties of are also investigated. Moreover, failure mechanisms under different conditions are elucidated to explain the failure behavior of the *CdTe* NWs.

## 2.Methodology

The uniaxial tension and compression simulations are performed for *CdTe* nanowire and only compression simulations are performed on nanopillar to check how ZB *CdTe* behaves under compression. MD simulations for characterizing mechanical behavior of *CdTe* nanowire is carried out using LAMMPS[33] software package and OVITO[34] is used for visualization of atomistic deformation processes. The SW (Stillinger-Weber )[35] potential is used to describe the interaction between *Cd* and *Te*. The aspect ratio of the nanowire height to width is kept constant as 10:1 for nanowire and 2:1 for nanopillar. Tension and compression are applied in crystal directions of [001], [110] and [111].

*CdTe* has a zinc blende (ZB) structure having lattice constant of 6.48 A°[36] . The NWs models are prepared by first creating a rectangular box of ZB *CdTe* with lattice constant a = 6.48 Å and later nanowires of specified diameter is cut from it with the help of atomsk tool[37] . The [111]-oriented ZB *CdTe* NWs are modeled with the x, y, and z axes oriented along the [$1\bar{2}1$], [$10\bar{1}$], and [111] directions, respectively. A few [001] and [110]-oriented *CdTe* NWs and nanopillar models are also developed to test the crystal orientation effect. The periodic boundary condition is maintained along the axis of the nanowire and nanopillar. The prepared [111]-oriented ZB *CdTe* nanowire model is shown in Fig. 1. Here, D and L stand for the diameter and length of the nanowire, respectively.

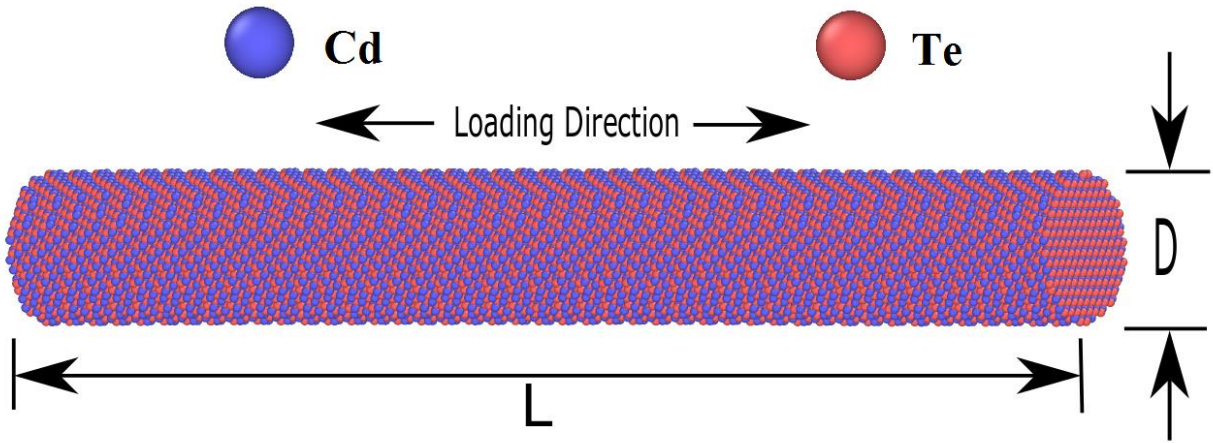

Figure 1: Schematic diagram of [111]-oriented ZB *CdTe* nanowire. The *Cd* and *Te* atoms are represented by blue and pink colors, respectively.

The type of structure, diameter, length of the nanowires and the total number of atoms calculated in each model are summarized in Table 1.

| Structure Type | Diameter, D(nm) | Length, L(nm) | Number of atoms |
| --- | --- | --- | --- |
| Nanowire | 2 | 20.2 | 1872 |
| | 3 | 29.5 | 6760 |
| | 4 | 39.8 | 14770 |
| | 5 | 49.4 | 28512 |
| | 6 | 60.6 | 50976 |
| Nanopillar | 10 | 20 | 46260 |

Before applying tensile and compressive load, the system energy is minimized using conjugate gradient algorithm. A constant integration time step of 1 fs is considered which is quite good for all the simulations. Before applying tensile and compressive load, constant NVE is performed for 10 ps. Then isothermal-isobaric (NPT) ensemble is applied for pressure equilibration at atmospheric pressure and prescribed temperature for 50 ps. Finally, the system is thermally equilibrated by canonical (NVT) ensemble for 50 ps. In order to control the temperature, a Nose-Hoover thermostat is employed in these steps. To equilibrate various state variables, the timesteps mentioned are chosen by trial and error for NVE, NPT and NVT simulation. Finally, the nanowire and nanopillar are deformed along their axis at a fixed strain rate of $10^9$ s$^{-1}$. This strain rate was successfully applied to predict the result in many tension and compression based MD simulations[38–40]. Both tensile and compressive test are conducted under NVT ensemble using a Nose-Hoover thermostat and carried out until the failure.

The simulation box is deformed uniaxially to calculate the atomic stress for obtaining stress-strain behavior. In our simulations, the atomic stresses are calculated using Virial stress theorem [38]. The equation of stress stands as

$$\sigma_{virial}(r) = \frac{1}{\Omega}\sum_i \left[\left(-m_i \dot{u}_i \otimes \dot{u}_i + \frac{1}{2}\sum_{j \neq i} r_{ij} \otimes f_{ij}\right)\right] \quad (1)$$

where the summation is done over all the atoms occupying the total volume, the mass of atom is represented by $m_i$ and displacement by $\dot{u}_i$. The relative position vector of atom is $r_{ij}$, the cross product is $\otimes$ and the interatomic force applied on atom $i$ by atom $j$ is $f_{ij}$.

## 3. Method validation

To validate the SW (Stillinger-Weber) potential employed in this study, the Young's modulus of [111]-oriented nanowires, lattice constants, and the cohesive energy of bulk *CdTe* are calculated. For Young's modulus a nanowire having diameter of 6nm at 100K is considered while for lattice constant and cohesive energy a cube of 2.592 nm × 2.592 nm × 2.592 nm of bulk *CdTe* is used. In this Study, Young's modulus is calculated from the stress-strain graphs using linear regression. Data obtained from the present study and data from the available literatures are presented in Table 2 for comparison. It is found that the calculated lattice constants, cohesive energy and Young's modulus agree well with other numerical and experimental studies.

|  | Lattice Constants(A°) | Cohesive Energy | Young's modulus (GPa) |
| --- | --- | --- | --- |
| Our calculation | 6.50 | 2.061 | 49(NW), 42(Bulk) |
| Literature Values (Simulations) | 6.486–6.573[41–43] | 2.068[36] | 49.55[44] |
| Literature Values (Experimental) | 6.486[45] | 2.060[46] | -- |

## 4. Results and Discussion

### 4.1. Effects of Temperature and Size

The stress-strain curve of [111]-oriented 4nm *CdTe* NWs under uniaxial tension and compression at strain rate $10^9$ s$^{-1}$ is shown in Fig. 2(a) for different temperatures. It is observed from the Fig. 2(a) that at 100K, the ultimate strength is about 7.7GPa with a failure strain of about 20.1% in case of tension while for compression the ultimate strength is about 3.01GPa with a failure strain of about 7.1%. Sharp fall of stress in Fig. 2(a) indicates brittle type failure of the material. It is also observed that as the temperature is increased from 100K to 600K, the fracture strength decreases from 7.7 GPa to 4.3 GPa in case of tension while for compression it decreases from 3.01 GPa to 1.6 GPa. This type of softening behavior at higher temperature is also observed for nanowires of other sizes in this study.

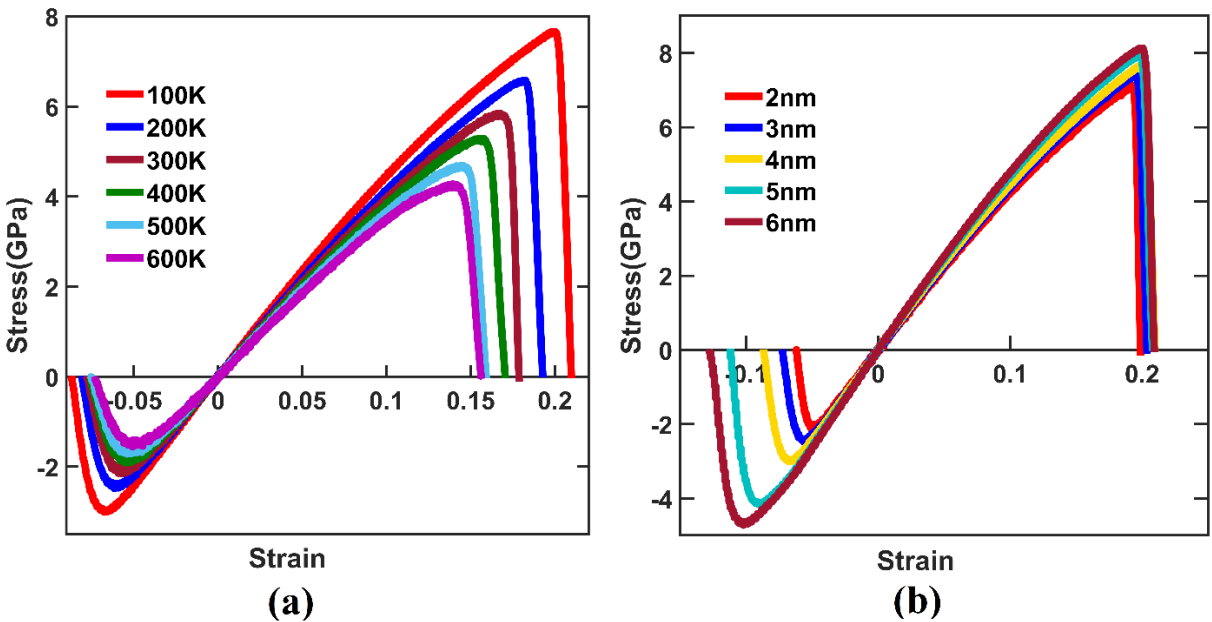

Figure 2: (a) Stress-strain curves for [111] oriented *CdTe* NWs at different temperature varied from 100K to 600K for a diameter of 4 nm and (b) Stress-strain curves for [111] oriented *CdTe* NWs of different diameters for temperature of 100K, and strain rate of $10^9$ s$^{-1}$.

Figure 2(b) shows stress–strain curve for [111]-oriented *CdTe* NWs with different diameters at 100K temperature, exhibiting the size effect. The diameter of the nanowires, in this case, is varied from 2 nm to 6 nm. The length of the nanowires is varied from 20.2 nm to 60.6 nm respectively to maintain the constant length to width ratio of 10:1. Here, the applied strain is $10^9$ s$^{-1}$ and the temperature is 100K. It is observed from the graph that all of the stress-strain curves for tension follow almost the same path until fracture under tension. However, for the length of nanowires used, the results show that in compression, the stress-strain curve largely depends on the diameter of the NWs. The larger the diameter, the larger the strength and elastic modulus have been observed. Similar kind of size dependent behavior for SiC was also found in the literature[27].

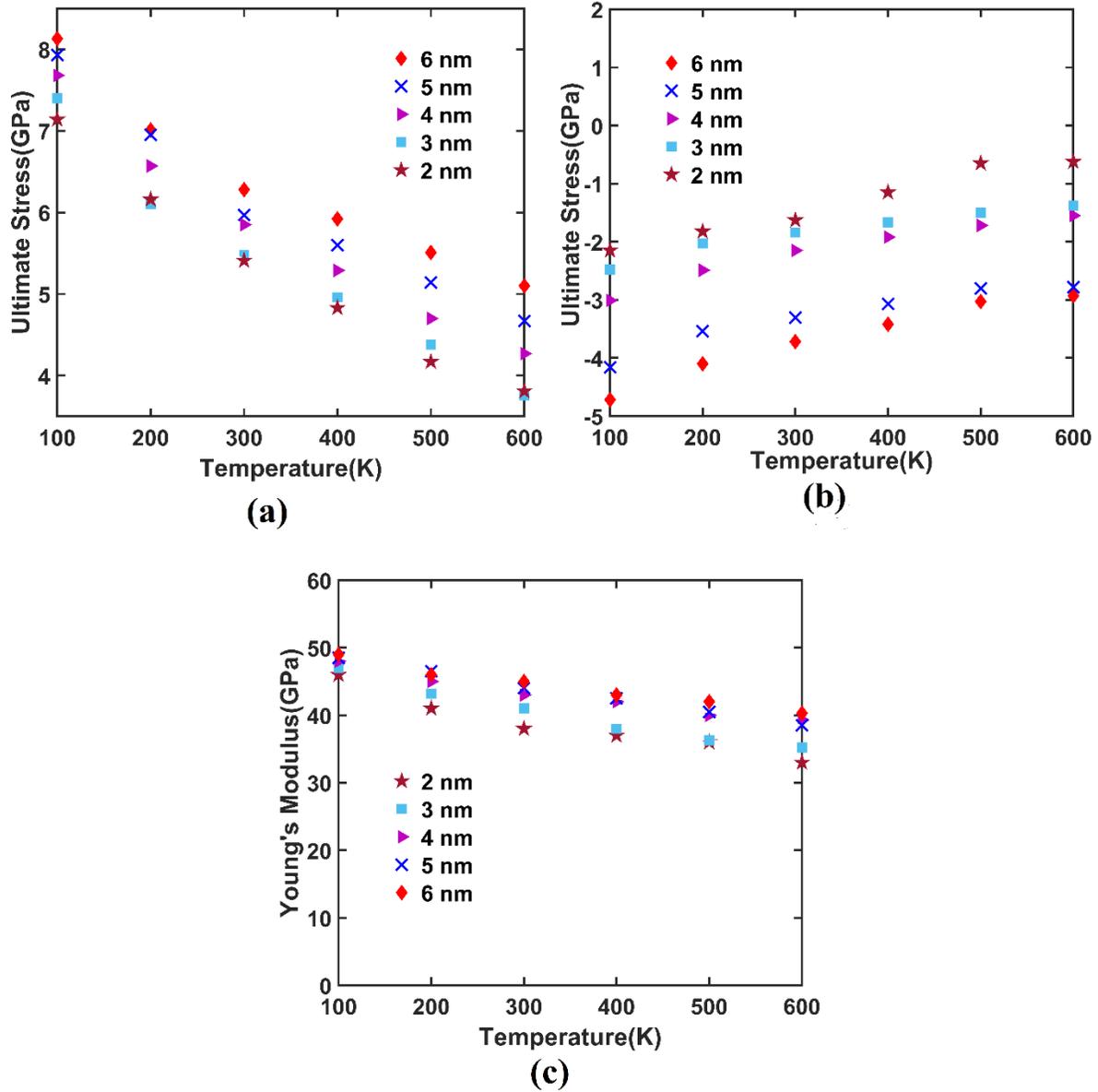

Figure 3: Variation of ultimate stresses of [111]-oriented *CdTe* in (a) tension and (b) compression with temperature, and for different diameters. (c) Variation of Young's modulus of ZB *CdTe* nanowires with temperatures, and for different diameters.

The variation of ultimate strength with temperature and size, in case of tension in Fig. 3(a) and in case of compression in Fig. 3(b) are elucidated. Figure shows that the ultimate strength decreases linearly with temperature. Moreover, the impact of size is more prominent in case of compression than in case of tension. The possibility for bonds to reach the critical bond length condition increases with temperatures which leads to bond breaking. The fracture is initiated imminently when a bond breaks because *CdTe* is a brittle type material. On the contrary, the crystal structure of ZB *CdTe* remains perfect at a lower temperature subjected to little or no excitation due to temperature, therefore, resulting high tensile stress.

Figure 3(c) shows the Young's modulus of [111]-oriented *CdTe* NWs as a function of NW diameter at different temperature ranging from 100K to 600K. It is observed that the Young's modulus decreases with temperature. Moreover, it decreases at a higher rate up to 400K, after that it reduces at a slower rate. At low temperatures, the structure behaves like perfect crystal that leads to this behavior.

**4.2. Effects of crystal orientation**

Figure 4(a) shows the stress–strain curves of *CdTe* with 4 nm diameter and of 300K temperature with [111], [110] and [001] crystal orientations. Under tension, the maximum fracture strength is 5.85 GPa for the [111] NW and the minimum strength has been obtained as 3.83GPa for the [001]-oriented NW. The [110]-orientation yields a fracture strength of 4.82 GPa.

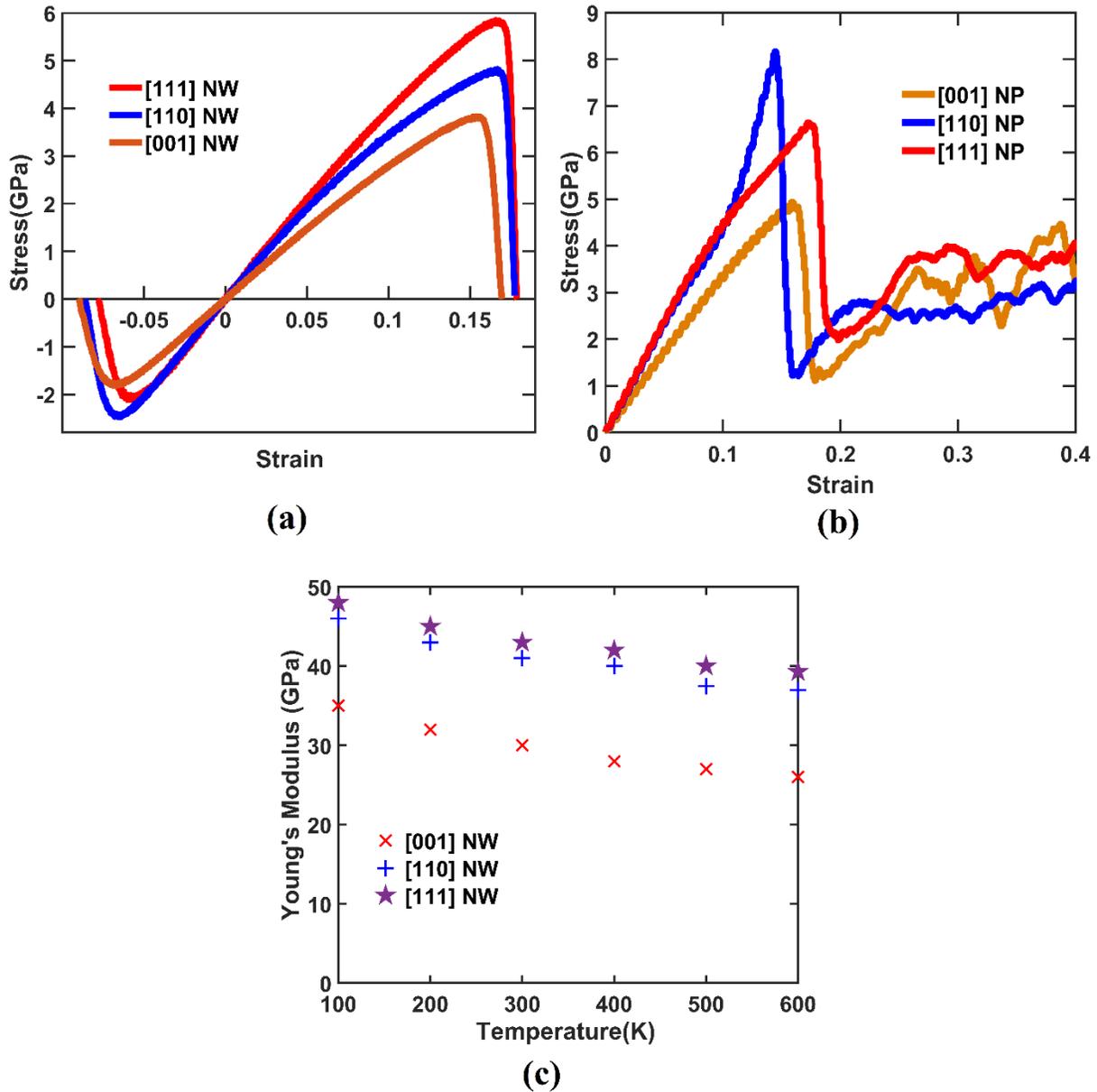

Figure 4: Stress–strain curves of [111], [110], and [001]-oriented (a) NWs of D = 4 nm at T = 300 K under both tension and compression. (b) NanoPillars of D=6 nm at 100k under compression. (c) Variation of Young's modulus of ZB *CdTe* nanowires with temperature, and for different crystals orientation of 4nm nanowires.

For diameters range 2–6 nm and at room temperature, it is observed that the fracture strength of the [111] orientation is the highest and for [001] orientation it is the lowest. Under tension, similar kind of orientation behavior was found for Si NWs [26]. However, in case of compression, the fracture strength is highest for the [110] direction (2.47 GPa) and it is the lowest for the [001] direction (1.81 GPa). In order to better observe how this ZB *CdTe* material behaves under compression, compressive load is given to 6 nm nanopillar (NP) having different crystal orientations. Similar kind of behavior is found in case of NP also. It is clear from the Fig. 4(b) that [111] NP has less compressive strength than [110] NP. While in case of [001], its compressive strength is lower than the other two. The bond breaking primarily depends on surface polarity, atomic coordination number, and bond length. Figure 5(a) shows the atomic arrangement of [001] oriented ZB *CdTe*, the bonds are arranged in a way that all the bonds make 45˚ with the loading direction (indicated by the arrow). As a result, these bonds fail at a lower compressive load. That is why [001] oriented structures have lower strength than the other two structures in both tension and compression. On the other hand, Fig. 5(b) and 5(c) shows that [110] and [111] oriented structure's atoms are singly bonded to the opposite surface or atomic layer. In both cases, the bonds are arranged in a way that all the bonds make 90˚ with the loading direction. Therefore, they are capable of taking higher tensile and compressive loads than the [001] oriented ZB *CdTe* structures. Furthermore, the spacing between atoms in [110] oriented structure is shorter than the spacing of [111] oriented structures and each atom is surrounded by opposite type atom resulting higher compressive strength. On the contrary, in [111] oriented ZB *CdTe*, there are two types of polarities in the atomic layers. From Fig. 5(c), it is evident that all atoms in a layer are positively charged (*Cd*) and all atoms in the adjacent layer are negatively charged (*Te*). As a result, there is an electrostatic attraction between two adjacent planes which makes it difficult to separate when tension is applied along the [111] orientation. However, under compression, when the layers come closer to each other, atoms with same polarity start to repel each other. Upon applying sufficient amount of load, the repelling force becomes significantly high in magnitude. At higher load, due to having larger atomic layer spacing, compressive load and repelling force, the effective compressive load increases. As a result, the [111]-oriented ZB *CdTe* structure have lower compressive strength than [110]-oriented ZB *CdTe* structure.

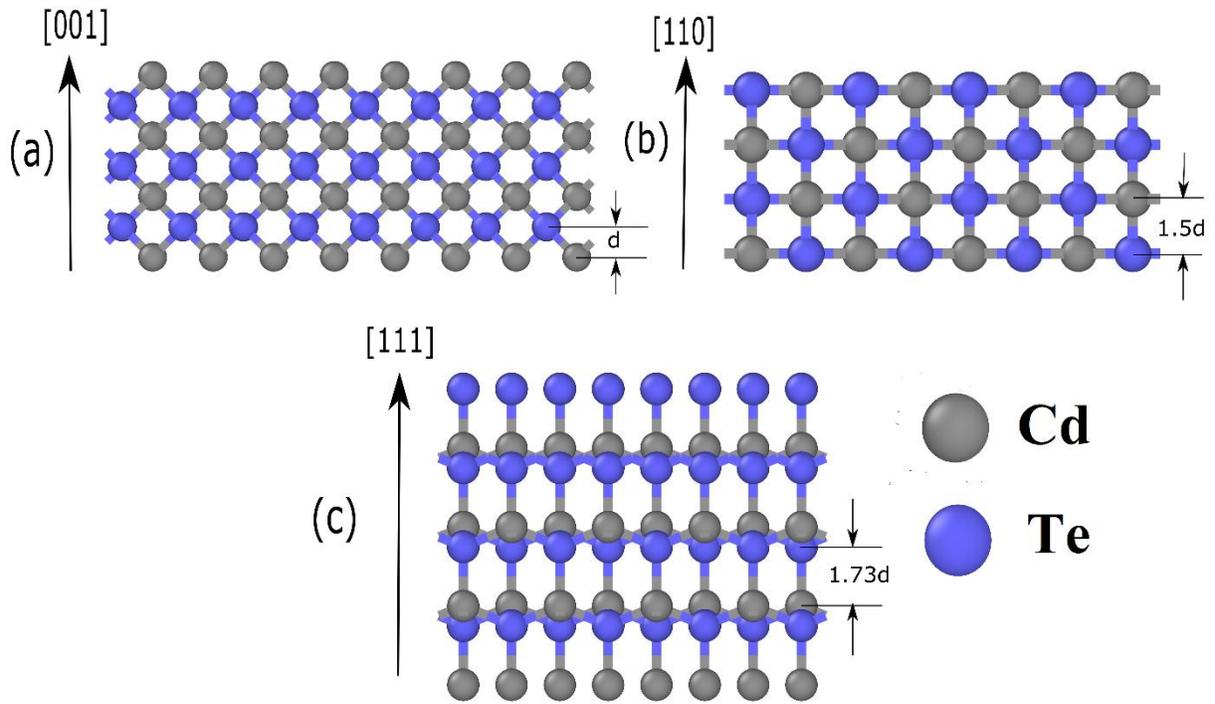

Figure 5: Representation of the atomic arrangements of (a) [001], (b) [110], (c) [111] oriented ZB *CdTe*, respectively. Here, ash and blue colored atoms represent *Cd* and *Te*, respectively.

Figure 4(c) represents the Young's modulus of *CdTe* NWs as a function of NWs crystal orientations at different temperature. The Young's modulus of [001], [110] and [111] oriented

*CdTe* NWs are obtained by fitting the stress–strain curve to a straight line. It is observed that the [111]-oriented *CdTe* NW has the highest value of elastic modulus. Moreover, the results are close for [111] and [110]-oriented NWs while for [001]-oriented NWs, the values are significantly lower than that of those two directions. Furthermore, the stress-strain curve has two distinct regions in the Figure which is similar to the study performed by Healy, et al. for *Fe* nanopillar [47]. The stress first increases linearly up to a certain point (yielding point) and then start to fluctuate as flow stress. It is observed that the average flow stress varies with crystal orientation. The maximum average flow stress is 3.76 GPa for the [111]-oriented NP and the minimum average flow stress for the [001]-oriented NP is 2.83 GPa. The [110]-orientation yields an average flow stress of 3.24 GPa. It is also observed that the average flow stress for the [111] direction is the highest and for [001] CdTe, it is the lowest. In case of tension, the flow stress is showing very little variation while in case of compression, the fluctuation can be observed significant due to the activation of several cross-slip system in the material.

### 4.3. Failure Mechanism

The [111]-oriented ZB *CdTe* NWs in the present study fails by creating void in [10-1] direction regardless of temperature and NW size. In Fig. 6, the failure phenomenon of the nanowire with diameter of 6 nm is shown at 300K temperature with the aid of construct surface mesh (CSM). At 15.74% strain, a void is initiated at NW surface which is marked with a red square in Fig. 6(a). It is found that loading in z direction causes very little plasticity in the nanowire and the fracture happens in a short strain region, from 15.74% to 16.71%. It is observed that the bonds are broken in [1-21] direction and void which is created due to bond breaking propagates in [10-1] direction not reported for Zinc Blende structure in previous literature, as illustrated in Fig. 6.

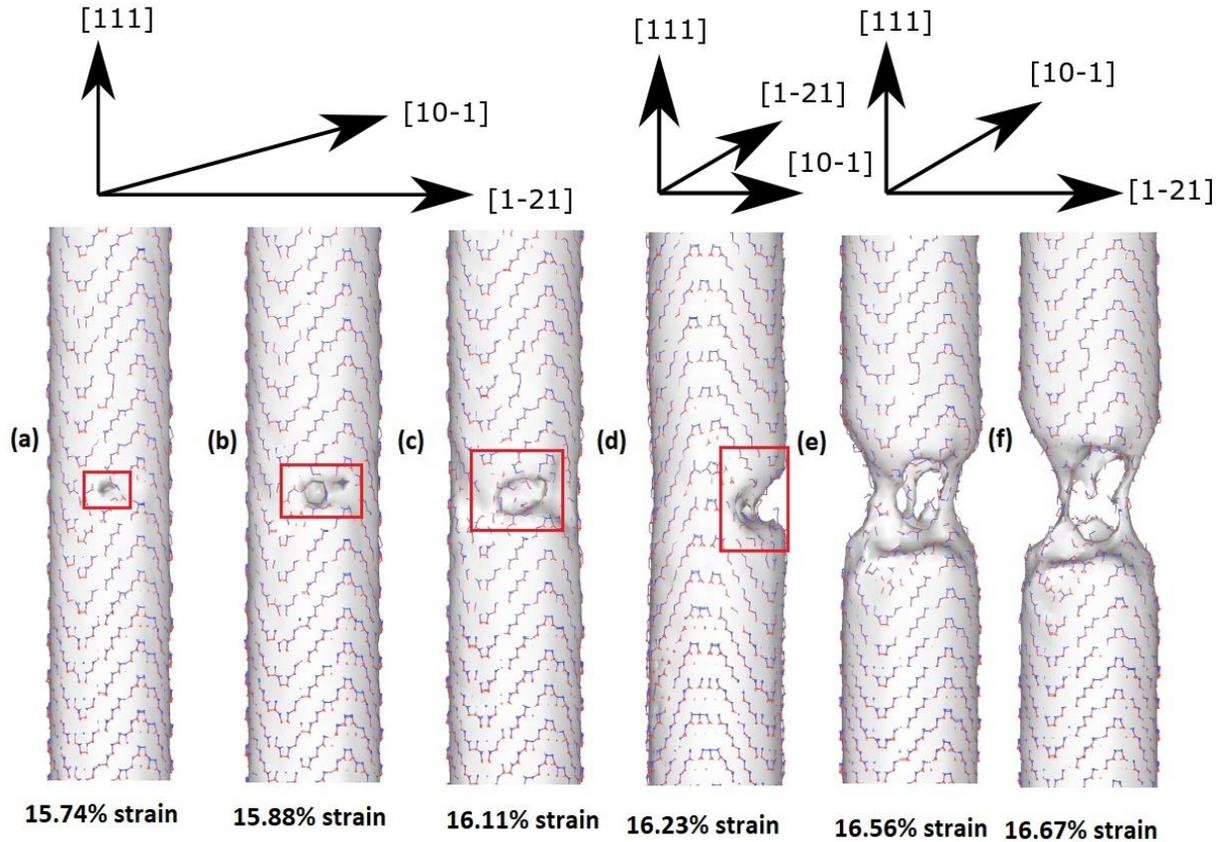

Figure 6: Failure mechanism of [111] oriented ZB *CdTe* NW with a diameter of 6nm at 300K. Voids are visualized using construct surface mesh (CSM). The red marked region in (a) indicates crack initiation and in (b), (c), and (d) indicates propagation of void. (a)-(c), (d), and (e)-(f) correspond to the three different crystal directions as shown on the top of the Figure.

To understand failure mechanism profoundly, atomic arrangement analysis of [111]-oriented ZB *CdTe* NWs is used which is shown in Fig 7. Breaking bonds along [10-1] direction requires high amount of energy because of its shorter bond gaps. Therefore, bond breaking is not seen in this [10-1] direction. On the other hand, along [1-21] direction gap between bonds is greater. Therefore, it would take less energy to break bonds along [1-21] direction. Bond breaking is a gradual process and it passes from one layer to another layer. When breaking of bonds in one layer is completed, bond breaking in another layer is initiated. So, in this way void propagates in [10-1] direction (see Fig. 6(d) in red marked region). Thus, fast rate of void propagation in [10-1] direction causes fracture in a short strain region.

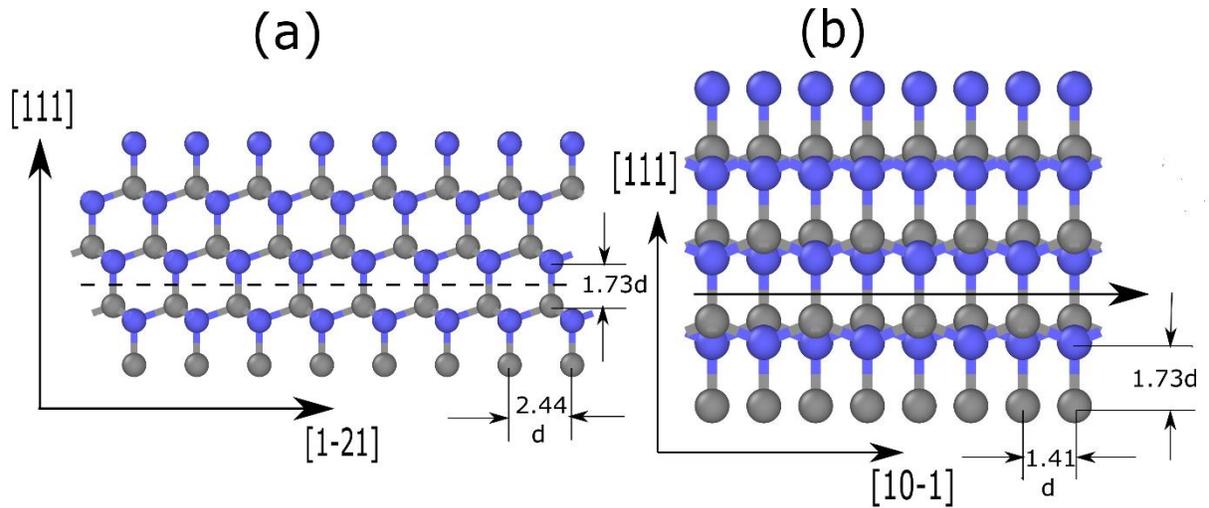

Figure 7: Representation of the atomic arrangements of (a) front view and (b) side view of [111] oriented ZB *CdTe*, respectively. Dotted line indicates those bonds that will break if crack nucleates and arrow indicates the void propagation direction. Here, ash and blue colored atoms represent *Cd* and *Te*, respectively.

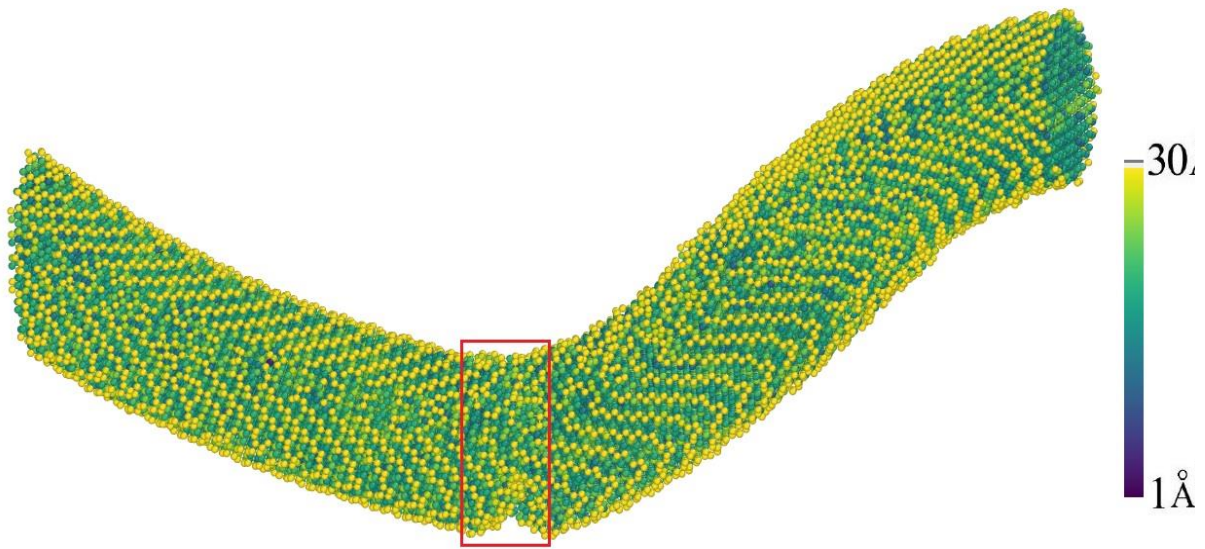

Figure 8: Deformation of [111] oriented ZB *CdTe* nanowire with 6nm diameter at 100K under compression and at a strain rate of $10^9$ s$^{-1}$. Colors of atoms follow the calculated values of the CSP. The red marked region indicates the location of defects formation as a result of buckling.

For compression, [111]-oriented ZB *CdTe* NW with diameter of 6 nm and temperature of 100K exhibits mechanical behavior that is different than that of tension. For a strain rate of -$10^9$ s$^{-1}$ the structure deforms and buckles and eventually fractures the nanowire which is shown in Fig. 8. Similar results have been obtained for NWs of other diameters also.

## 4.4. Effects of Strain rates

Finally, the influence of strain rate on the mechanical properties of *CdTe* NWs is depicted in Fig. 9. Stress–strain curves of a [111]-oriented *CdTe* NW with 5 nm diameter size at 300K are shown for a strain rate range from $10^8$ s$^{-1}$ to $10^{11}$ s$^{-1}$. It is observed from the Figure that for both uniaxial tension and compression, the fracture strength and the strain reduce with the decrement of strain rate. This type of phenomenon occurs because defects can nucleate easily at lower stresses if more time is given. The curves of both tension and compression for the NWs above 5×10$^9$ are distorted due to the disorderliness and fluctuation of atoms stemming from the impact of such unusual high strain. Similar kind of strain dependent behavior was observed in case of SiC[27].

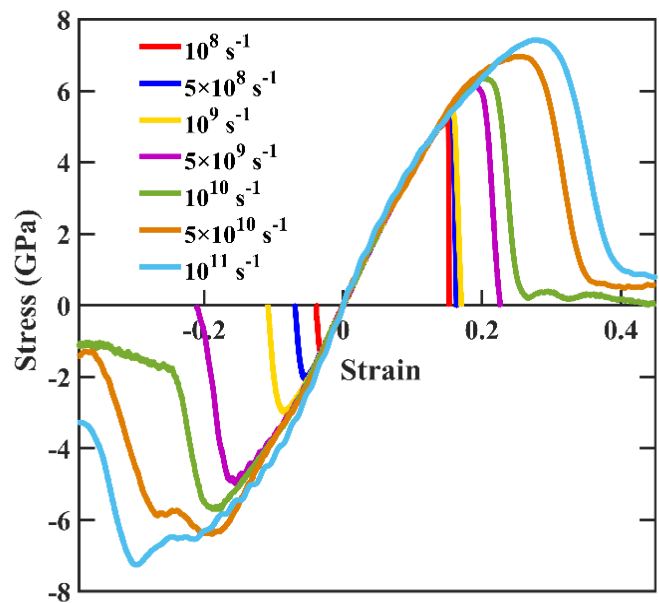

Figure 9: Stress–strain curves for [111]-oriented ZB *CdTe* nanowires for tensile and compressive loading at different strain rates for 5nm diameter and 300K temperature.

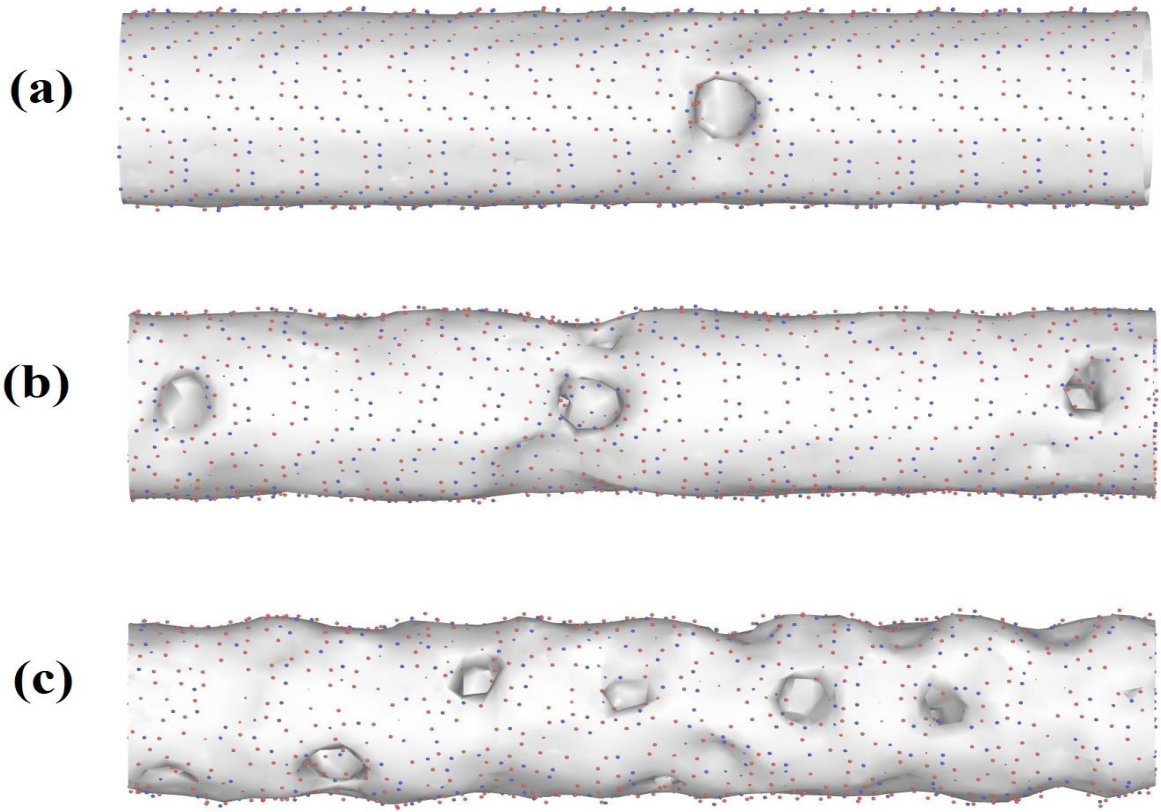

Figure 10: Deformation of [111] oriented ZB *CdTe* nanowire for increasing tensile strain rates. Voids are visualized using construct surface mesh (CSM). (a) single void occurs under the strain rate of $10^9$ s$^{-1}$. (b) Few voids are nucleated for the intermediate rate of $10^{10}$ s$^{-1}$. (c) Cascade of voids in the nanowire at the highest rate of $10^{11}$ s$^{-1}$.

The failure mechanisms of [111]-oriented ZB *CdTe* NW for the three tensile strain rates $10^9$ s$^{-1}$, $10^{10}$ s$^{-1}$ and $10^{11}$ s$^{-1}$ are illustrated in Fig. 10. The change in failure behavior for different strain rates is presented in these Figures. In Fig. 10(a), for slowest strain rate, only one void is created leading to a brittle fracture. For intermediate strain rate of $10^{10}$ s$^{-1}$, more voids are observed to cause the failure (see Fig. 10(b)). For the highest strain rate of $10^{11}$ s$^{-1}$, cascade of voids are nucleated throughout the [111] oriented ZB *CdTe* NW structure. (Fig. 10(c))

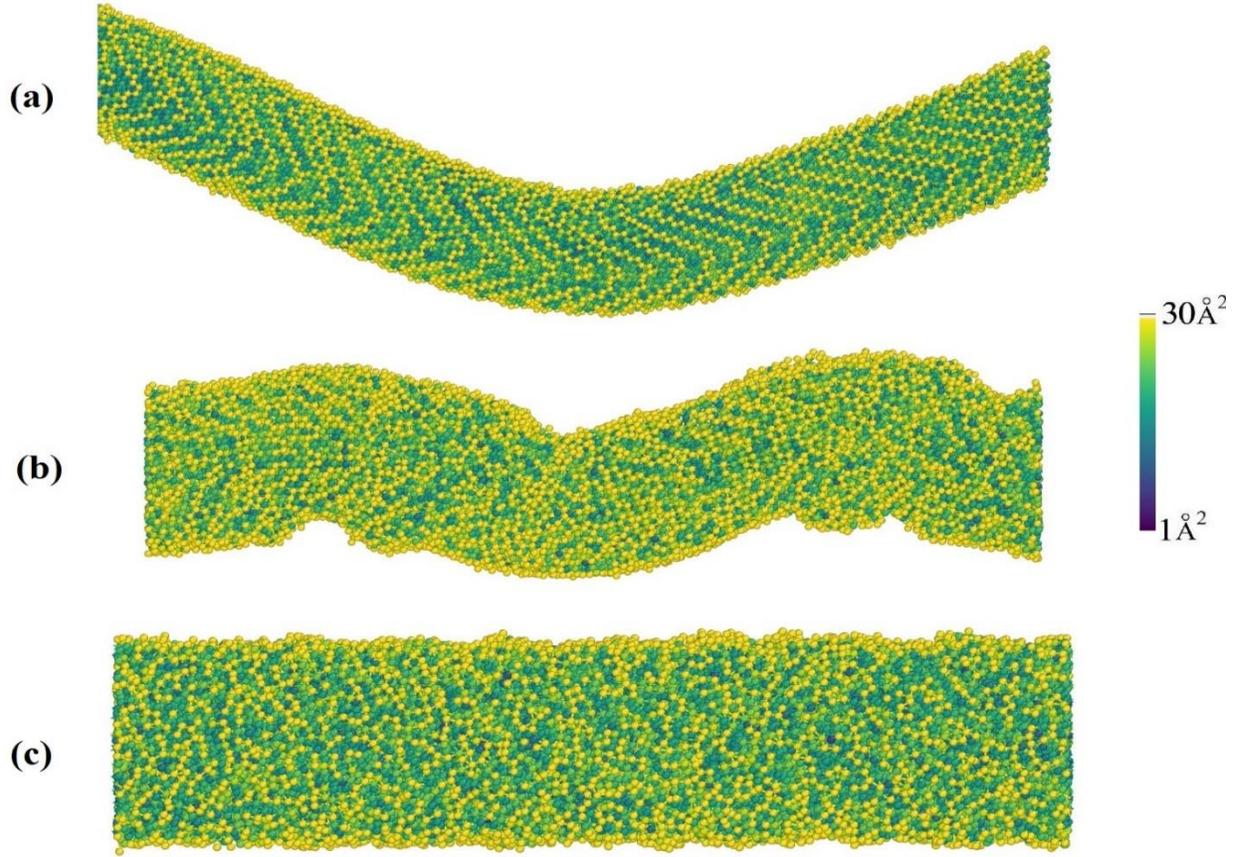

Figure 11: Deformation of [111] oriented ZB *CdTe* nanowire with 5nm diameter at 300K temperature for increasing compressive strain rates. Colors of atoms follow the calculated values of the CSP. (*a*) Buckling behavior under the strain rate of -$10^9$ s$^{-1}$. (*b*) Multiple buckling and plastic deformations under the intermediate strain rate of -$10^{10}$ s$^{-1}$. (*c*) Homogeneous amorphization under the highest strain rate of -$10^{11}$ s$^{-1}$.

The mechanical behavior of [111]-oriented ZB *CdTe* NWS under different compressive strain rates are also distinctive. For gradual increase of compressive strain rates there is a clear transition in the [111]-oriented *CdTe* NWs behavior from single buckling to multiple buckling to homogeneous amorphization. This behavior is illustrated in Fig. 11. At the low strain rate of $10^9$ s$^{-1}$, the nanowires develop a single buckling that leads to fracture of the nanowire. However, at the intermediate rate of $10^{10}$ s$^{-1}$, multiple buckling is generated along the NWs accompanied by irreversible plastic deformations, as illustrated in Fig. 11(b). At the highest compressive strain rate of $10^{11}$ s$^{-1}$ NW is homogeneously amorphized. This behavior is consistent with some previous studies for different nanowires[27].

## 5.Conclusion

In this investigation, Zinc blende Cadmium Telluride nanowires are studied by molecular dynamics simulations. Here, atomistic simulations have been carried out to investigate both the tensile and compressive mechanical behavior of these nanowires considering different sizes, temperatures, crystal orientation and strain rates. Both ultimate strength and Young's modulus show an inverse relation with temperature under both tension and compression. It is also demonstrated that size has negligible effect on the tensile behavior but during compression the ultimate strength and elastic modulus are diameter dependent. However, the NWs of *CdTe* under tension shows considerably higher strength than NWs under compression. One of the main findings of this study is that fracture strength of the [111]-oriented *CdTe* NWs is always higher than that of the [110] -oriented NWs under tension, while in compression, the fracture strength of the [111] NWs is always lower than that of the [110] NWs. The [111]-oriented ZB *CdTe* NWs fails by creating void in [10-1] direction regardless of temperature and NW size which is a new failure mechanism. Investigation suggests that these phenomena of ZB *CdTe* nanowires are controlled by bond length, atomic spacing and electrostatic forces. Finally, it is observed that with increasing the strain rate, both the ultimate strength and strain increase. The failure mechanism for low to high strain rates are also elucidated. This investigation provides a comprehensive understanding on temperature, size, crystal orientation, and strain rate dependent mechanical properties and fracture phenomenon of ZB *CdTe* NWs which has enumerable application in NEMS/MEMS.

## 6. Acknowledgements

The authors of this paper would like to thank Department of Mechanical Engineering, BUET for providing the computing resources and Multiscale Mechanical Modelling and Research Network (MMMRN) group of the same department for the technical support to conduct the research.

## 7.References

[1] Y. Huang, X. Duan, Y. Cui, L.J. Lauhon, K.-H. Kim, C.M. Lieber, Logic Gates and Computation from Assembled Nanowire Building Blocks, Science. 294 (2001) 1313–1317. doi:10.1126/science.1066192.

[2] X. Duan, Y. Huang, R. Agarwal, C.M. Lieber, Single-nanowire electrically driven lasers, Nature. 421 (2003) 241–245. doi:10.1038/nature01353.

[3] B. Tian, X. Zheng, T.J. Kempa, Y. Fang, N. Yu, G. Yu, J. Huang, C.M. Lieber, Coaxial silicon nanowires as solar cells and nanoelectronic power sources, Nature. 449 (2007) 885–889. doi:10.1038/nature06181.

[4] C.M. Lieber, Z.L. Wang, Functional Nanowires, MRS Bull. 32 (2007) 99–108. doi:10.1557/mrs2007.41.

[5] D.L. Dreifus, R.M. Kolbas, K.A. Harris, R.N. Bicknell, R.L. Harper, J.F. Schetzina, CdTe metal-semiconductor field-effect transistors, Appl. Phys. Lett. 51 (1987) 931–933. doi:10.1063/1.98805.

[6] D.L. Dreifus, R.M. Kolbas, J.R. Tassitino, R.L. Harper, R.N. Bicknell, J.F. Schetzina, Electrical properties of CdTe metal–semiconductor field effect transistors, J. Vac. Sci. Technol. A. 6 (1988) 2722–2724. doi:10.1116/1.575493.

[7] Y.-J. Lee, H.-J. Ryu, S.-W. Lee, S.-J. Park, H.-J. Kim, Comparison of ultra-high-resolution parallel-hole collimator materials based on the CdTe pixelated semiconductor SPECT system, Nucl. Instrum. Methods Phys. Res. Sect. Accel. Spectrometers Detect. Assoc. Equip. 713 (2013) 33–39. doi:10.1016/j.nima.2013.03.014.

[8] H. Tsutsui, T. Ohtsuchi, K. Ohmori, S. Baba, CdTe semiconductor X-ray imaging sensor and energy subtraction method using X-ray energy information, IEEE Trans. Nucl. Sci. 40 (1993) 95–101. doi:10.1109/23.212323.

[9] A.E. Rakhshani, Heterojunction properties of electrodeposited CdTe/CdS solar cells, J. Appl. Phys. 90 (2001) 4265–4271. doi:10.1063/1.1397279.

[10] J. Britt, C. Ferekides, Thin-film CdS/CdTe solar cell with 15.8% efficiency, Appl. Phys. Lett. 62 (1993) 2851–2852. doi:10.1063/1.109629.

[11] Y. Ye, L. Dai, T. Sun, L.P. You, R. Zhu, J.Y. Gao, R.M. Peng, D.P. Yu, G.G. Qin, High-quality CdTe nanowires: Synthesis, characterization, and application in photoresponse devices, J. Appl. Phys. 108 (2010) 044301. doi:10.1063/1.3474991.

[12] M. Shaygan, K. Davami, N. Kheirabi, C.K. Baek, G. Cuniberti, M. Meyyappan, J.-S. Lee, Single-crystalline CdTe nanowire field effect transistors as nanowire-based photodetector, Phys. Chem. Chem. Phys. 16 (2014) 22687–22693. doi:10.1039/C4CP03322A.

[13] X.D. Han, K. Zheng, Y.F. Zhang, X.N. Zhang, Z. Zhang, Z.L. Wang, Low-Temperature In Situ Large-Strain Plasticity of Silicon Nanowires, Adv. Mater. 19 (2007) 2112–2118. doi:10.1002/adma.200602705.

[14] F. Östlund, K. Rzepiejewska-Malyska, K. Leifer, L.M. Hale, Y. Tang, R. Ballarini, W.W. Gerberich, J. Michler, Brittle-to-Ductile Transition in Uniaxial Compression of Silicon Pillars at Room Temperature, Adv. Funct. Mater. 19 (2009) 2439–2444. doi:10.1002/adfm.200900418.

[15] F. Östlund, P.R. Howie, R. Ghisleni, S. Korte, K. Leifer, W.J. Clegg, J. Michler, Ductile–brittle transition in micropillar compression of GaAs at room temperature, Philos. Mag. 91 (2011) 1190–1199. doi:10.1080/14786435.2010.509286.

[16] T. Kizuka, Y. Takatani, K. Asaka, R. Yoshizaki, Measurements of the atomistic mechanics of single crystalline silicon wires of nanometer width, Phys. Rev. B. 72 (2005) 035333. doi:10.1103/PhysRevB.72.035333.

[17] R.A. Minamisawa, M.J. Süess, R. Spolenak, J. Faist, C. David, J. Gobrecht, K.K. Bourdelle, H. Sigg, Top-down fabricated silicon nanowires under tensile elastic strain up to 4.5%, Nat. Commun. 3 (2012) 1096. doi:10.1038/ncomms2102.

[18] J. Michler, K. Wasmer, S. Meier, F. Östlund, K. Leifer, Plastic deformation of gallium arsenide micropillars under uniaxial compression at room temperature, Appl. Phys. Lett. 90 (2007) 043123. doi:10.1063/1.2432277.

[19] G. Stan, S. Krylyuk, A.V. Davydov, R.F. Cook, Compressive Stress Effect on the Radial Elastic Modulus of Oxidized Si Nanowires, Nano Lett. 10 (2010) 2031–2037. doi:10.1021/nl100062n.

[20] A. San Paulo, J. Bokor, R.T. Howe, R. He, P. Yang, D. Gao, C. Carraro, R. Maboudian, Mechanical elasticity of single and double clamped silicon nanobeams fabricated by the vapor-liquid-solid method, Appl. Phys. Lett. 87 (2005) 053111. doi:10.1063/1.2008364.

[21] Y. Calahorra, O. Shtempluck, V. Kotchetkov, Y.E. Yaish, Young's Modulus, Residual Stress, and Crystal Orientation of Doubly Clamped Silicon Nanowire Beams, Nano Lett. 15 (2015) 2945–2950. doi:10.1021/nl5047939.

[22] S. Hoffmann, I. Utke, B. Moser, J. Michler, S.H. Christiansen, V. Schmidt, S. Senz, P. Werner, U. Gösele, C. Ballif, Measurement of the Bending Strength of Vapor−Liquid−Solid Grown Silicon Nanowires, Nano Lett. 6 (2006) 622–625. doi:10.1021/nl052223z.

[23] S.E. Root, S. Savagatrup, C.J. Pais, G. Arya, D.J. Lipomi, Predicting the Mechanical Properties of Organic Semiconductors Using Coarse-Grained Molecular Dynamics Simulations, Macromolecules. 49 (2016) 2886–2894. doi:10.1021/acs.macromol.6b00204.

[24] J. Guénolé, J. Godet, S. Brochard, Deformation of silicon nanowires studied by molecular dynamics simulations, Model. Simul. Mater. Sci. Eng. 19 (2011) 074003. doi:10.1088/0965-0393/19/7/074003.

[25] S.A. Dayeh, J. Wang, N. Li, J.Y. Huang, A.V. Gin, S.T. Picraux, Growth, Defect Formation, and Morphology Control of Germanium–Silicon Semiconductor Nanowire Heterostructures, Nano Lett. 11 (2011) 4200–4206. doi:10.1021/nl202126q.

[26] K. Kang, W. Cai, Size and temperature effects on the fracture mechanisms of silicon nanowires: Molecular dynamics simulations, Int. J. Plast. 26 (2010) 1387–1401. doi:10.1016/j.ijplas.2010.02.001.

[27] H. Tsuzuki, J.P. Rino, P.S. Branicio, Dynamic behaviour of silicon carbide nanowires under high and extreme strain rates: a molecular dynamics study, J. Phys. Appl. Phys. 44 (2011) 055405. doi:10.1088/0022-3727/44/5/055405.

[28] G. Cheng, T.-H. Chang, Q. Qin, H. Huang, Y. Zhu, Mechanical Properties of Silicon Carbide Nanowires: Effect of Size-Dependent Defect Density, Nano Lett. 14 (2014) 754–758. doi:10.1021/nl404058r.

[29] T.H. Pial, T. Rakib, S. Mojumder, M. Motalab, M.A.S. Akanda, Atomistic investigations on the mechanical properties and fracture mechanisms of indium phosphide nanowires, Phys. Chem. Chem. Phys. 20 (2018) 8647–8657. doi:10.1039/C7CP08252E.

[30] K. Kang, W. Cai, Brittle and ductile fracture of semiconductor nanowires – molecular dynamics simulations, Philos. Mag. 87 (2007) 2169–2189. doi:10.1080/14786430701222739.

[31] Z. Wang, X. Zu, L. Yang, F. Gao, W.J. Weber, Atomistic simulations of the size, orientation, and temperature dependence of tensile behavior in GaN nanowires, Phys. Rev. B. 76 (2007) 045310. doi:10.1103/PhysRevB.76.045310.

[32] L. Dai, W.C.D. Cheong, C.H. Sow, C.T. Lim, V.B.C. Tan, Molecular Dynamics Simulation of ZnO Nanowires: Size Effects, Defects, and Super Ductility, Langmuir. 26 (2010) 1165–1171. doi:10.1021/la9022739.

[33] S. Plimpton, Fast Parallel Algorithms for Short-Range Molecular Dynamics, J. Comput. Phys. 117 (1995) 1–19. doi:10.1006/jcph.1995.1039.

[34] A. Stukowski, Visualization and analysis of atomistic simulation data with OVITO–the Open Visualization Tool, Model. Simul. Mater. Sci. Eng. 18 (2010) 015012. doi:10.1088/0965-0393/18/1/015012.

[35] Z.Q. Wang, D. Stroud, A.J. Markworth, Monte Carlo study of the liquid CdTe surface, Phys. Rev. B. 40 (1989) 3129–3132. doi:10.1103/PhysRevB.40.3129.

[36] M.B. Kanoun, W. Sekkal, H. Aourag, G. Merad, Molecular-dynamics study of the structural, elastic and thermodynamic properties of cadmium telluride, Phys. Lett. A. 272 (2000) 113–118. doi:10.1016/S0375-9601(00)00403-5.

[37] P. Hirel, Atomsk: A tool for manipulating and converting atomic data files, Comput. Phys. Commun. 197 (2015) 212–219. doi:10.1016/j.cpc.2015.07.012.

[38] S. Saha, M. Abdul Motalab, M. Mahboob, Investigation on mechanical properties of polycrystalline W nanowire, Comput. Mater. Sci. 136 (2017) 52–59. doi:10.1016/j.commatsci.2017.04.025.

[39] S. Mojumder, Molecular dynamics study of plasticity in Al-Cu alloy nanopillar due to compressive loading, Phys. B Condens. Matter. 530 (2018) 86–89. doi:10.1016/j.physb.2017.10.119.

[40] S. Muhammad Nahid, S. Nahian, M. Motalab, T. Rakib, S. Mojumder, M. Mahbubul Islam, Tuning the mechanical properties of silicene nanosheet by auxiliary cracks: a molecular dynamics study, RSC Adv. 8 (2018) 30354–30365. doi:10.1039/C8RA04728F.

[41] N.E. Christensen, O.B. Christensen, Electronic structure of ZnTe and CdTe under pressure, Phys. Rev. B. 33 (1986) 4739–4746. doi:10.1103/PhysRevB.33.4739.

[42] K. Strössner, S. Ves, W. Dieterich, W. Gebhardt, M. Cardona, High pressure X-ray investigations of phase transitions in Cd1−xMnxTe, Solid State Commun. 56 (1985) 563–565. doi:10.1016/0038-1098(85)90955-X.

[43] M.L. Cohen, Calculation of bulk moduli of diamond and zinc-blende solids, Phys. Rev. B. 32 (1985) 7988–7991. doi:10.1103/PhysRevB.32.7988.

[44] E. Deligoz, K. Colakoglu, Y. Ciftci, Elastic, electronic, and lattice dynamical properties of CdS, CdSe, and CdTe, Phys. B Condens. Matter. 373 (2006) 124–130. doi:10.1016/j.physb.2005.11.099.

[45] J.M. Rowe, R.M. Nicklow, D.L. Price, K. Zanio, Lattice dynamics of cadmium telluride, Phys. Rev. B. 10 (1974) 671–675. doi:10.1103/PhysRevB.10.671.

[46] W.A. Harrison, Electronic Structure and the Properties of Solids: The Physics of the Chemical Bond, Courier Corporation, 2012.

[47] C.J. Healy, G.J. Ackland, Molecular dynamics simulations of compression–tension asymmetry in plasticity of Fe nanopillars, Acta Mater. 70 (2014) 105–112. doi:10.1016/j.actamat.2014.02.021.